\documentclass[pra,twocolumn]{revtex4}

\usepackage{type1cm}
\usepackage[dvipdfm]{graphicx}

\usepackage{amsmath,amssymb,bm,amsthm}
\def\beq{\begin{equation}}
\def\eeq{\end{equation}}
\def\nbeq{\begin{equation*}}
\def\neeq{\end{equation*}}
\def\<{\langle}
\def\>{\rangle}

\begin{document}
\title{Natural correlation between system and thermal reservoir}
\author{Takashi Mori}
\email{
mori@spin.phys.s.u-tokyo.ac.jp}
\affiliation{
Department of Physics, Graduate School of Science,
University of Tokyo, Bunkyo-ku, Tokyo 113-0033, Japan
}

\begin{abstract}
Non-Markovian corrections to the Markovian quantum master equation of an open quantum system are investigated 
up to the second order of the interaction between the system of interest and a thermal reservoir.
The concept of ``natural correlation'' is discussed.
When the system is naturally correlated with a thermal reservoir, the time evolution of the reduced density matrix looks Markovian even in a short-time regime.
If the total system was initially in an ``unnatural'' state, the natural correlation is established during the time evolution,
and after that the time evolution becomes Markovian in a long-time regime.
It is also shown that for a certain set of reduced density matrices, the naturally correlated state does not exist.
If the initial reduced density matrix has no naturally correlated state, the time evolution is inevitably non-Markovian in a short-time regime.
\end{abstract}
\maketitle

The time evolution of the reduced density matrix $\rho_{\rm S}(t)$ of the system of interest in contact with a large thermal reservoir 
is described by the quantum master equation~\cite{Breuer_text,Weiss_text,Kubo_text}.
When the coupling constant $\lambda$ between the system and the reservoir is very small,
the time scale of the relaxation is well separated from the time scale of the microscopic motion of the system and that of the reservoir,
and hence the Markov approximation is justified.
Indeed, the Markovian quantum master equation is rigorously derived in the van Hove limit, 
$\lambda\rightarrow 0$ with $\lambda^2 t$ fixed~\cite{Davies_text}.
However, in some real situations, such as atoms in photonic band gap media~\cite{John1995,Quang1997,Vats1998,Chen2012}
and an optical cavity coupled to a structured reservoir~\cite{Longhi2006},
the deviation from the van Hove limit is important and the non-Markov effect is not negligible.

In this paper, the non-Markovian corrections are investigated up to the order of $\lambda^2$.
The key concept discussed in this work is the {\it natural correlation} between the system of interest and the thermal reservoir.
The following properties are to be clarified.
When the total system starts from a {\it naturally correlated state}, 
the time evolution of the system of interest obeys the Markovian quantum master equation even if there is deviation from the van Hove limit.
On the other hand, if the initial state is not a naturally correlated state, the time evolution must be non-Markovian in short times $t\lesssim\tau_{\rm R}$,
where $\tau_{\rm R}$ is a characteristic time of the motion of the thermal reservoir.
Even in this case, after a sufficiently long time $t\gg\tau_{\rm R}$, the time evolution approximately becomes Markovian
since the natural correlation is established during the time evolution.

The further observation of this work is that there is a set ${\cal U}$ of reduced density matrices such that for any $\rho_{\rm S}\in{\cal U}$,
the naturally correlated state does {\it not} exist.
Therefore, if some $\rho_{\rm S}\in{\cal U}$ is chosen as an initial state of the system of interest,
the time evolution is inevitably non-Markovian in the short time regime $t\lesssim\tau_{\rm R}$.
After some time $t$, $\rho_{\rm S}(t)$ will get out of ${\cal U}$ and the natural correlation will develop there.

Now the setup is explained.
The Hamiltonian of the total system is given by $H_{\rm T}=H_{\rm S}+H_{\rm R}+\lambda H_{\rm I}$,
where $H_{\rm S}$ is the Hamiltonian of the system of interest, $H_{\rm R}$ is the Hamiltonian of the thermal reservoir,
and $H_{\rm I}$ is the interaction Hamiltonian.
The coupling constant $\lambda$ is relatively small but not vanishingly small, and hence we consider the non-Markovian corrections up to $O(\lambda^2)$.
The density matrix of a whole system is denoted by $\rho_{\rm T}(t)$ 
and the reduced density matrix is then given by $\rho_{\rm S}(t)={\rm Tr}_{\rm R}\rho_{\rm T}(t)$.
The initial state of the reservoir is assumed to be close to an equilibrium state 
$\rho_{\rm R}^{\rm eq}=e^{-\beta H_{\rm R}}/{\rm Tr}_{\rm R}e^{-\beta H_{\rm R}}$ where $\beta>0$ is the inverse temperature of the reservoir.
At an initial time $t_0$, it is assumed that $\lim_{\lambda\rightarrow 0}\rho_{\rm T}(t_0)=\rho_{\rm S}(t_0)\rho_{\rm R}^{\rm eq}$.
Without loss of generality, it is assumed that ${\rm Tr}_{\rm R}H_{\rm I}\rho_{\rm R}^{\rm eq}=0$. 
By applying the standard Nakajima-Zwanzig projection operator method~\cite{Nakajima1958,Zwanzig1960,Kubo_text},
the time evolution of $\rho_{\rm S}(t)$ for $t\geq t_0$ is given by
\begin{align}
\dot{\rho}_{\rm S}(t)=&-iL_{\rm S}\rho_{\rm S}(t)
-\lambda^2\int_{t_0}^tdt'{\rm Tr}_{\rm R}e^{-(t-t')QiL_{\rm T}}QL_{\rm I}\rho_{\rm R}^{\rm eq}\rho_{\rm S}(t')
\nonumber \\
&-i\lambda{\rm Tr}_{\rm R}L_{\rm I}e^{-(t-t_0)QiL_{\rm T}}Q\rho_{\rm T}(t_0),
\label{eq:gQME}
\end{align}
which is formally exact.
Here the projection operators $P$ and $Q=1-P$ are defined as $P:=\rho_{\rm R}^{\rm eq}{\rm Tr}_{\rm R}$.
The Liouvillian $L_{X}$ ($X=$T, S, R, I) is defined as $L_X(\cdot):=[H_X,(\cdot)]$.
Up to $O(\lambda^2)$, the second term of the right-hand side of Eq.~(\ref{eq:gQME}) is simplified as $-\Lambda_0\rho_{\rm S}(t)+\Lambda_{t-t_0}\rho_{\rm S}(t)$ with
\beq
\Lambda_t:=\lambda^2\int_t^{\infty}dt'{\rm Tr}_{\rm R}L_{\rm I}L_{\rm I}(-t'),
\label{eq:Lambda}
\eeq
where $L_{\rm I}(t)(\cdot)=[H_{\rm I}(t),(\cdot)]$
in the interaction picture, $H_{\rm I}(t)=e^{it(H_{\rm S}+H_{\rm R})}H_{\rm I}e^{-it(H_{\rm S}+H_{\rm R})}$.
If the terms up to $O(\lambda^2)$ of the last term of Eq.~(\ref{eq:gQME}) are denoted by ${\cal I}_{t-t_0}\rho_{\rm T}(t_0)$, we have
\beq
\dot{\rho}_{\rm S}(t)=-(iL_{\rm S}+\Lambda_0)\rho_{\rm S}(t)+\Lambda_{t-t_0}\rho_{\rm S}(t)+{\cal I}_{t-t_0}\rho_{\rm T}(t_0).
\label{eq:QME}
\eeq
In the Markov approximation, we take the limit of $t_0\rightarrow-\infty$.
As a result, the following Markovian quantum master equation (the Redfield equation~\cite{Redfield1957}) is obtained:
\beq
\dot{\rho}_{\rm S}^{\rm (M)}(t)=-(iL_{\rm S}+\Lambda_0)\rho_{\rm S}^{\rm (M)}(t).
\label{eq:Markov}
\eeq
In this paper we call $\Lambda_{t-t_0}\rho_{\rm S}(t)$ and ${\cal I}_{t-t_0}\rho_{\rm T}(t_0)$ in Eq.~(\ref{eq:QME}) non-Markovian contributions.
In particular, the former is referred to as the ``non-Markovian relaxation term'' and the latter is referred to as the ``initial correlation term''.
These terms are negligible in the van Hove limit, but here we consider finite corrections up to $O(\lambda^2)$.
It is noted that the initial correlation is often omitted by considering a product initial state, but here we do not make such an assumption.

Up to $O(\lambda^2)$,
we can evaluate the non-Markovian contributions by the perturbation theory.
The result is
\begin{align}
\rho_{\rm S}(t)=e^{-(t-t_0)(iL_{\rm S}+\Lambda_0)}
\{\rho_{\rm S}(t_0)+\delta\rho_1[t-t_0;\rho_{\rm S}(t_0)]
\nonumber \\
+\delta\rho_2[t-t_0;\rho_{\rm T}(t_0)]\},
\label{eq:solution}
\end{align}
where
\begin{align}
\delta\rho_1[t;\rho_{\rm S}]&:=\int_0^tdt'e^{iL_{\rm S}t'}\Lambda_{t'}e^{-iL_{\rm S}t'}\rho_{\rm S}
\nonumber \\
=&\lambda^2\int_0^tdt'\int_0^{\infty}dt''
{\rm Tr}_{\rm R}L_{\rm I}(t')L_{\rm I}(-t'')\rho_{\rm R}^{\rm eq}\rho_{\rm S},
\label{eq:delta1}
\\
\delta\rho_2[t;\rho_{\rm T}]&:=\int_0^tdt'e^{iL_{\rm S}t'}{\cal I}_{t'}\rho_{\rm T}.
\label{eq:delta2}
\end{align}
The non-Markovian corrections are expressed by $\delta\rho_1[t-t_0;\rho_{\rm S}(t_0)]$ and $\delta\rho_2[t-t_0;\rho_{\rm T}(t_0)]$.
Here let us write the interaction Hamiltonian as $H_{\rm I}=\sum_iX_iY_i$, 
where $X_i$ and $Y_i$ are operators acting on the Hilbert space of the system of interest and the reservoir, respectively.
We define the correlation functions of the reservoir as $C_{ij}(t):={\rm Tr}_{\rm R}\rho_{\rm R}^{\rm eq}Y_i(t)Y_j(0)$
and the characteristic decay time of $\{C_{ij}(t)\}$ is denoted by $\tau_{\rm R}$.
Each correction, $\delta\rho_1[t-t_0;\rho_{\rm S}(t_0)]$ and $\delta\rho_2[t-t_0;\rho_{\rm T}(t_0)]$, 
then converges to some constant matrix for $t-t_0\gg\tau_{\rm R}$.
Therefore, when $t-t_0\gg\tau_{\rm R}$, the non-Markovian contribution is taken into account 
as a slippage of the initial condition of the Markovian quantum master equation,
\beq
\rho_{\rm S}^{(M)}(t_0)=\rho_{\rm S}(t_0)+\delta\rho_1[\infty;\rho_{\rm S}(t_0)]+\delta\rho_2[\infty;\rho_{\rm T}(t_0)].
\eeq
Su\'arez, Silbey, and Oppenheim~\cite{Suarez1992} originally pointed out that 
an initial condition of the Markovian quantum master equation slips due to the non-Markovian effect.
Gaspard and Nagaoka~\cite{Gaspard-Nagaoka1999} analyzed this slippage of the initial condition 
and derived the same expression of $\delta\rho_1[\infty;\rho_{\rm S}(t_0)]$, but they did not consider the contribution from the initial correlation, $\delta\rho_2$.
As for calculations of equilibrium fluctuations, van Kampen took into account the effect of the initial correlation~\cite{van_Kampen_text,van_Kampen2004}.
In the present work, we shall show that a new insight can be obtained 
if {\it both} the non-Markovian relaxation term and the initial correlation term are treated in a unified way for {\it general} initial conditions 
(not restricted to equilibrium states or product states).

Let us consider a time $t_1$ with $t_1-t_0\gg\tau_{\rm R}$.
In this case, $\delta\rho_i[t-t_0;\cdot]\simeq\delta\rho_i[\infty;\cdot]$ ($i=1,2$) 
and $\rho_{\rm S}(t)$ approximately obeys the Markovian quantum master equation~(\ref{eq:Markov}) for $t\geq t_1$.
Now let us choose $t_1$ as a new initial time.
Then $\rho_{\rm S}(t)=e^{-(t-t_1)\{ iL_{\rm S}+\Lambda_0)}(\rho_{\rm S}(t_1)
+\delta\rho_1[t-t_1;\rho_{\rm S}(t_1)]+\delta\rho_2[t-t_1;\rho_{\rm T}(t_1)]\}$ for $t\geq t_1$.
On the other hand, as was explained above, $\rho_{\rm S}(t)$ should obey the Markovian quantum master equation for $t>t_1$ 
if we regard $t_0$ as an initial time.
Since the time evolution of $\rho_{\rm S}(t)$ should be independent of our choice of an initial time,
the above observation yields that $\delta\rho_1[t-t_1;\rho_{\rm S}(t_1)]+\delta\rho_2[t-t_1;\rho_{\rm T}(t_1)]$ is very small for $t>t_1$.

In general, $\delta\rho_1[t-t_1;\rho_{\rm S}(t_1)]+\delta\rho_2[t-t_1;\rho_{\rm T}(t_1)]$ is not exactly equal to zero as long as $t_1-t_0$ is finite.
However, if
\beq
\delta\rho_1[t-t_1;\rho_{\rm S}(t_1)]+\delta\rho_2[t-t_1;\rho_{\rm T}(t_1)]=0 \quad \text{for any }t\geq t_1
\label{eq:condition}
\eeq
is satisfied, the time evolution of the system of interest is Markovian for $t\geq t_1$ exactly up to $O(\lambda^2)$.

When $\rho_{\rm T}$ satisfies Eq.~(\ref{eq:condition}),
we say that the system is {\it naturally correlated} with a thermal reservoir.
More precisely, for a given $\rho_{\rm S}$, we define the {\it naturally correlated state} $\rho_{\rm T}^{\rm (N)}$
as a matrix satisfying (i) $\rho_{\rm T}^{\rm (N)}\geq 0$, (ii) ${\rm Tr}_{\rm R}\rho_{\rm T}^{\rm (N)}=\rho_{\rm S}$,
(iii) $\lim_{\lambda\rightarrow 0}Q\rho_{\rm T}^{\rm (N)}=0$, and
(iv) $\delta\rho_1[t;\rho_{\rm S}]+\delta\rho_2[t;\rho_{\rm T}^{\rm (N)}]=0$ for any $t\geq 0$.
The state $\rho_{\rm T}^{\rm (N)}$ satisfying the above conditions is a special one,
but this special correlation is automatically generated during the time evolution even if we do not prepare a naturally correlated initial state.
In this sense, $\rho_{\rm T}^{\rm (N)}$ is ``natural''.

For example, if the initial state is given by a product state $\rho_{\rm T}=\rho_{\rm S}\rho_{\rm R}^{\rm eq}$,
Eq.~(\ref{eq:condition}) does not hold unless $\delta\rho_1[t;\rho_{\rm S}]=0$ for all $t>0$, which is generally not the case.
In this sense, for a given $\rho_{\rm S}$, the product form $\rho_{\rm T}=\rho_{\rm S}\rho_{\rm R}^{\rm eq}$ is ``unnatural''.
A trivial example of naturally correlated states is an equilibrium state of the total system,
$\rho_{\rm T}^{\rm eq}=e^{-\beta H_{\rm T}}/{\rm Tr}_{\rm SR}e^{-\beta H_{\rm T}}$.
We can show $\delta\rho_1+\delta\rho_2=0$ for $\rho_{\rm T}^{\rm eq}$; see also Ref.~\cite{Mori2008}.
It is stressed that it is highly nontrivial whether nonequilibrium naturally correlated states exist.
The condition (iv) is required for all $t>0$ and that is a very strong condition.

In this paper, as for the naturally correlated states, the following two results are obtained.
The first result is that for any fixed $\rho_{\rm S}$, the conditions (ii), (iii), and (iv) are satisfied if and only if
\beq
\rho_{\rm T}^{\rm (N)}=\rho_{\rm S}\rho_{\rm R}^{\rm eq}+i\lambda\int_0^{\infty}dtL_{\rm I}(-t)\rho_{\rm S}\rho_{\rm R}^{\rm eq}+\lambda D.
\label{eq:natural}
\eeq
This result is obtained by Eqs.~(\ref{eq:delta1}) and (\ref{eq:delta2}).
Here, $D$ is an operator satisfying ${\rm Tr}_{\rm R}D=0$ and ${\rm Tr}_{\rm R}L_{\rm I}(t)D=0$ for all $t$~\footnote{
For a sufficiently complex reservoir, it is expected that only $D=0$ is allowed.
In this case, the reservoir is called ``ergodic''. 
The boson bath given by Eq.~(\ref{eq:spin-boson}) is not ergodic in this sense
although non-linear couplings between boson modes might make the reservoir ergodic.}.
If Eq.~(\ref{eq:natural}) satisfies condition (i), it gives an explicit expression of the naturally correlated state for a given $\rho_{\rm S}$.
Because $\rho_{\rm T}^{\rm (N)}\geq 0$ for $\lambda\rightarrow 0$, it is expected that $\rho_{\rm T}^{(N)}\geq 0$ at least for small values of $\lambda$. 
The second result, however, insists that for any fixed $\lambda\neq 0$, there is a set ${\cal U}$ such that 
$\rho_{\rm T}^{\rm (N)}$ given by Eq.~(\ref{eq:natural}) does not satisfy condition (i) for any $\rho_{\rm S}\in{\cal U}$;
there is no naturally correlated state for $\rho_{\rm S}\in{\cal U}$.
Therefore, if the initial state of the system of interest is chosen from ${\cal U}$,
the time evolution is inevitably non-Markovian in a short-time regime $t\lesssim\tau_{\rm R}$.
After a sufficiently long time $t\gg\tau_{\rm R}$, $\rho_{\rm S}(t)$ will get out of ${\cal U}$ 
and the density matrix of the total system will approach a naturally correlated state.

According to the Pechukas theorem~\cite{Pechukas1994}, 
a linear assignment of $\rho_{\rm S}\rightarrow\rho_{\rm T}\geq 0$ for all $\rho_{\rm S}\geq 0$ is impossible 
except for the product assignment $\rho_{\rm S}\rightarrow\rho_{\rm S}\rho_{\rm R}$.
If only $D=0$ were allowed, Eq.~(\ref{eq:natural}) would give a linear assignment, and hence this assignment would not work for all $\rho_{\rm S}$;
${\cal U}$ is not empty.
In general, however, $D\neq 0$ is allowed.
In that case, $D$ may depend on $\rho_{\rm S}$ nonlinearly, and the Pechukas theorem does not ensure that ${\cal U}$ is not empty.
Our second result shows that ${\cal U}$ is not empty even in that case.

So far we have assumed that a characteristic correlation time $\tau_{\rm R}$ exists.
However, it is remarked that even if $\{C_{ij}(t)\}$ decays algebraically and thus there is no characteristic time $\tau_{\rm R}$,
the above results are valid as long as $\delta\rho_1[t;\rho_{\rm S}]$ has the limit of $t\rightarrow\infty$.
If this limit does not exist, the non-Markovian effect cannot be treated perturbatively.
This point will be explained later.

Although we have defined the set ${\cal U}$, it is hard to judge whether a given $\rho_{\rm S}$ is in ${\cal U}$.
Therefore, we explicitly construct a subset ${\cal U}'\subset{\cal U}$, which is much easier to calculate numerically.
The argument relies on the variational consideration.
If $\rho_{\rm T}^{\rm (N)}$ is positive semidefinite, $\<\Psi|\rho_{\rm T}^{\rm (N)}|\Psi\>\geq 0$ for any state vector $|\Psi\>$.
Here we restrict state vectors into the following set of trial state vectors $\{|\Psi_{\alpha}\>\}_{\alpha}$:
\beq
|\Psi_{\alpha}\>=|\phi_0^{\rm (S)}\>|\psi_{\alpha}^{\rm (R)}\>
-i\lambda\xi\int_0^tdt'H_{\rm I}(t')|\phi_0^{\rm (S)}\>|\psi_{\alpha}^{\rm (R)}\>,
\eeq
where $\xi\in\mathbb{R}$, $t\geq 0$, $|\phi_0^{\rm (S)}\>$ is the eigenvector of $\rho_{\rm S}$ with the smallest eigenvalue $p_0$,
and $|\psi_{\alpha}^{\rm (R)}\>$ is an energy eigenstate of the reservoir, $H_{\rm R}|\psi_{\alpha}^{\rm (R)}\>=E_{\alpha}|\psi_{\alpha}^{\rm (R)}\>$.
For $\xi=1$, the above form of $|\Psi_{\alpha}\>$ corresponds to the time-evolved quantum state 
obtained by the first order perturbation theory in the interaction picture starting from a product initial state.
Because the natural correlation is established during the time evolution of the total system, this choice of trial state vectors seems to be natural.
Here $\xi$ and $t$ play the role of variational parameters.

Here we consider the quantity $\sum_{\alpha}\<\Psi_{\alpha}|\rho_{\rm T}^{\rm (N)}|\Psi_{\alpha}\>$, which is independent of $D$.
Substituting Eq.~(\ref{eq:natural}), we obtain
\beq
\sum_{\alpha}\<\Psi_{\alpha}|\rho_{\rm T}^{\rm (N)}|\Psi_{\alpha}\>=p_0+\lambda^2[\xi^2A(t)-\xi B(t)],
\eeq
where
\beq
A(t):=\int_0^tdt'\int_0^tdt''\<\phi_0^{\rm (S)}|{\rm Tr}_{\rm R}H_{\rm I}(t')\rho_{\rm S}\rho_{\rm R}^{\rm eq}H_{\rm I}(t'')|\phi_0^{\rm (S)}\>,
\eeq
which is non-negative $A(t)\geq 0$ for $t\geq 0$,
and
\beq
\lambda^2B(t):=\<\phi_0^{\rm (S)}|\delta\rho_1[t;\rho_{\rm S}]|\phi_0^{\rm (S)}\>.
\eeq
Since $\xi\in\mathbb{R}$ and $t\geq 0$ are arbitrary, we choose them so that $\sum_{\alpha}\<\Psi_{\alpha}|\rho_{\rm T}^{\rm (N)}|\Psi_{\alpha}\>$
becomes minimum.
Putting $\xi=B(t)/2A(t)$ and minimizing with respect to $t>0$, we have
\beq
\sum_{\alpha}\<\Psi_{\alpha}|\rho_{\rm T}^{\rm (N)}|\Psi_{\alpha}\>=p_0-\lambda^2\sup_{t>0}\frac{B(t)^2}{4A(t)},
\label{eq:U'}
\eeq
which is strictly lower than $p_0$.
If the right-hand side of Eq.~(\ref{eq:U'}) is negative, it contradicts the non-negativity of $\rho_{\rm T}^{\rm (N)}$.
Hence, we define ${\cal U}'$ as 
\beq
{\cal U}':=\left\{\rho_{\rm S}:p_0-\lambda^2\sup_{t>0}\frac{B(t)^2}{4A(t)}<0\right\}.
\label{eq:def_U'}
\eeq
Clearly ${\cal U}'\subset{\cal U}$ and it is not an empty set.
When the system of interest is in a pure state, $p_0=0$ and thus the right-hand side of Eq.~(\ref{eq:U'}) is negative.
That is, all the pure states do not have their naturally correlated states.
It is because the density matrix of the total system must be a product state $\rho_{\rm S}\rho_{\rm R}$ in this case
and there is no room for building up any correlation between the system of interest and the thermal reservoir.

\begin{figure}[tb]
\begin{center}
\includegraphics[width=6cm]{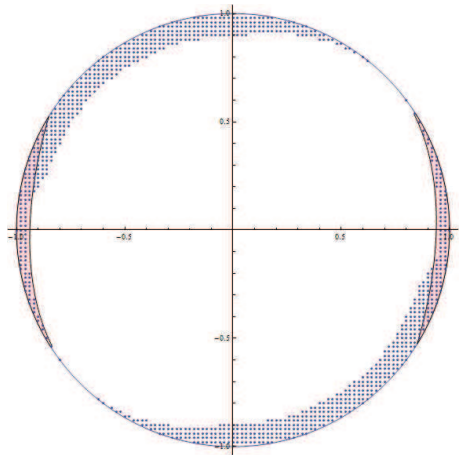}
\caption{(Color online) The regions of ${\cal U}'$ (dotted region) and ${\cal N}$ (shaded region).
Up to $O(\lambda^2)$, ${\cal N}\subset{\cal U}'$.
All the physical states are in the unit circle (solid line).}
\label{fig:region}
\end{center}
\end{figure} 

As a demonstration, let us consider the spin-boson model, in which a two level system is in contact with an infinitely large boson bath.
The Hamiltonian is given by 
\beq
H_{\rm S}=\varepsilon S^z, \quad H_{\rm R}=\sum_r\omega_rb_r^{\dagger}b_r, \quad H_{\rm I}=X_1Y_1
\label{eq:spin-boson}
\eeq
with $X_1=S^x$ and $Y_1=\sum_r\nu_r(b_r^{\dagger}+b_r)$.
Here $b_r$ and $b_r^{\dagger}$ are the annihilation and the creation operators of the boson, $[b_r,b_{r'}^{\dagger}]=\delta_{r,r'}$ and
$[b_r,b_{r'}]=[b_r^{\dagger},b_{r'}^{\dagger}]=0$.
The vector $\bm{S}$ denotes the spin-1/2 operator.
The spectral density of the boson bath is assumed to be the ohmic one with the Lorentz-Drude cutoff~\cite{Breuer_text},
$J(\omega):=\sum_r\nu_r^2\delta(\omega_r-\omega)=\omega\Omega^2/(\omega^2+\Omega^2)$,
for which we can numerically evaluate $A(t)$, $B(t)$, and also the right-hand side of Eq.~(\ref{eq:U'}).
The reduced density matrix is parametrized by $x$, $y$, and $z$ with $\sqrt{x^2+y^2+z^2}\leq 1$
so that $\rho_{\rm S}=(1/2)\hat{I}+xS^x+yS^y+zS^z$,
where $\hat{I}$ is the identity matrix.
In this Bloch-sphere representation, $p_0=(1-\sqrt{x^2+y^2+z^2})/2$.

The region of ${\cal U}'$ is shown for $z=0$ on the $xy$-plane in Fig.~\ref{fig:region}.
In this figure, the parameters are set to $\varepsilon=\Omega=\beta=1$ and $\lambda=0.5$.
This parameter choice is just for demonstration of the result.
The domain of ${\cal U}'$ is not so large; the length of this domain from the surface of the Bloch sphere is proportional to $\lambda^2$.
Although we cannot obtain the region of ${\cal U}$ explicitly, we expect that this ${\cal U}'$ gives a good approximation of ${\cal U}$.
Figure~\ref{fig:region} implies that naturally correlated states exist for a large part of physically allowed reduced density matrices $\{\rho_{\rm S}\}$.

It is a well known problem that the Markovian quantum master equation~(\ref{eq:Markov}) sometimes violates the non-negativity of the reduced density matrix.
It depends on the initial state $\rho_{\rm S}$ whether or not the non-negativity is violated during the Markovian time evolution.
In Fig.~\ref{fig:region}, we also show the domain ${\cal N}$ of initial states $\{\rho_{\rm S}\}$
for which the non-negativity of $\{\rho_{\rm S}^{\rm (M)}(t)\}_{t\geq 0}$ with $\rho_{\rm S}^{\rm (M)}(0)=\rho_{\rm S}$ is violated.
We numerically find ${\cal N}\subset{\cal U}'$ up to $O(\lambda^2)$;
all the initial conditions for which the non-negativity is violated during the Markovian time evolution do not have their naturally correlated states~\footnote
{Strictly speaking, ${\cal N}$ is not completely a subset of ${\cal U}'$ for a finite value of $\lambda$.
However, a typical length of the region ${\cal N}\backslash{\cal U}'$ shrinks faster than $\lambda^2$.
The statement ``${\cal N}\subset{\cal U}'$ up to $O(\lambda^2)$'' should be interpreted in this sense.}.
The violation of non-negativity is due to the improper choice of an initial state of the Markovian equation.
This statement was already indicated by the previous work~\cite{Suarez1992} and the present study gives a theoretical ground for this statement.
Moreover, the fact that ${\cal U}'$ is larger than ${\cal N}$ means that we should take special care to use the Markovian quantum master equation 
even if the Markovian time evolution does not violate the non-negativity and therefore it looks physical.
The general criterion for making use of the Markovian master equation is whether $\rho_{\rm S}(0)$ is in ${\cal U}$ or not.
For $\rho_{\rm S}\in{\cal U}$, the time evolution cannot be Markovian in a short-time regime.
Therefore, an initial state of the Markovian quantum master equation must be chosen from the outside of ${\cal U}$.

As was mentioned just before presenting the main results, our perturbative treatment of the non-Markovian effect is invalid for the case 
where $\delta\rho_1[t;\rho_{\rm S}]$ does not have the limit of $t\rightarrow\infty$.
This limitation excludes the case where an ohmic or sub-ohmic Bson bath~\cite{Weiss_text}, i.e. $J(\omega)\sim\omega^s$ with $s\leq 1$ at low frequencies,
is at zero temperature and, moreover, $\{X_i\}$ have non-zero diagonal elements in the basis diagonalizing $H_{\rm S}$.
In the spin-boson model calculated above, $X_1=S^x$ and it does not have diagonal elements in the basis diagonalizing $S^z$,
and hence we can safely treat the non-Markovian effect perturbatively.
However, if we consider another coupling with $X_2=S^z$, 
the perturbative treatment of the non-Markovian effect becomes invalid for an ohmic or sub-ohmic reservoir at zero temperature.
In that case, very long time correlations of the boson bath, which originate from the singularity around $\omega=0$,
cause the breakdown of the perturbative treatment of the non-Markovian effect.

Even if the perturbation analysis works well, of course, there are very small Markovian and non-Markovian effects coming from higher-order terms than $\lambda^2$.
Higher-order terms might become important for very long times.
For instance, the terms of $O(\lambda^4)$ may change the stationary state in $O(\lambda^2)$~\cite{Mori2008,Fleming2011}.
However, they do not play important roles for the transient relaxation dynamics, which is the major interest of this work.
 
In conclusion, it has been argued that there are special states called the ``naturally correlated states'' 
by the evaluation of the non-Markovian contributions up to $O(\lambda^2)$,
and moreover it has been shown that there is no naturally correlated state for some reduced density matrices $\rho_{\rm S}\in{\cal U}$.
If the system is in a naturally correlated state, the time evolution is Markovian up to $O(\lambda^2)$ regardless of the deviation from the van Hove limit.
On the other hand, if the system is not in a naturally correlated state, the later time evolution is non-Markovian 
until the natural correlation develops for $\rho_{\rm S}(t)$ being outside of ${\cal U}$.
The subset ${\cal U}'$ of ${\cal U}$ is given by Eq.~(\ref{eq:def_U'}), which allows explicit numerical computation.
The domain of ${\cal U}'$ is demonstrated for the spin-boson model.
As a result, it is numerically shown that the domain of ${\cal U}'$ is wider than that of ${\cal N}$,
in which the Markovian time evolution violates the non-negativity of the density matrix.
It confirms the previous qualitative argument given in Ref.~\cite{Suarez1992} 
on the violation of non-negativity due to the neglected non-Markovian effect in the short-time regime.
It is made clear that the underlying reason why the non-Markovian effect is important for some initial conditions
is the impossibility of developing the natural correlation between the system and the thermal reservoir at short times. 

The author thanks Seiji Miyashita for helpful discussions during the early stages of this work
and Tomotaka Kuwahara for valuable comments.
This work was supported by the Sumitomo foundation.

\end{document}